\documentclass{aa501}

\usepackage{graphics}

\begin{document}

\font\small=cmr8           
\font\petit=cmcsc10        

\font\bbf=cmbx10 scaled\magstep1 
\font\bbbf=cmbx10 scaled\magstep2 
\font\bbbbf=cmbx10 scaled\magstep3 

\def\subti#1{\par\vskip0.8cm{\bf\noindent #1}\par\vskip0.4cm}
\def\ti#1{\par\vskip1.6cm{\bbf\noindent #1}\par\vskip0.8cm}
\def\bigti#1{\par\vfil\eject{\bbbf\noindent #1}\par\vskip1.6cm}

\def\cbigti#1{\par\vskip2.0cm{\bbbf\noindent\centerline
{#1}\par\vskip0.5cm}}
\def\cti#1{\par\vskip1.0cm{\bbf\noindent\centerline
{#1}\par\vskip0.4cm}}
\def\csubti#1{\par\vskip0.5cm{\bf\noindent\centerline
{#1}\par\vskip0.3cm}}

\def\doublespace {\baselineskip 22pt}        

\def\eqd{\buildrel \rm d \over =}    
\def\p{\partial}           
\def\px{\partial _x}           
\def\py{\partial _y}           
\def\pz{\partial _z}           
\def\pt{\partial _t}           
\def\ssum{\textstyle\sum}
\def\arr{\rightarrow}
\def\id{\equiv}
\def\eqv{\leftrightarrow}
\def\fol{\rightarrow}
\let\prop=\sim
\def\gapprox{\;\rlap{\lower 2.5pt            
 \hbox{$\sim$}}\raise 1.5pt\hbox{$>$}\;}       
\def\lapprox{\;\rlap{\lower 2.5pt            
 \hbox{$\sim$}}\raise 1.5pt\hbox{$<$}\;} 

\def\ang{\,{\rm\AA}}
\def\cm{\,{\rm cm}}
\def\km{\,{\rm km}}
\def\kpc{\,{\rm kpc}}
\def\second{\,{\rm sec}}     
\def\erg{\,{\rm erg}}
\def\ev{\,{\rm e\kern-.1em V}}
\def\kev{\,{\rm ke\kern-.1em V}}
\def\k{\,{\rm K}}
\def\K{\,{\rm K}}
\def\gauss{\,{\rm gauss}}
\def\SFU{\,{\rm SFU}}

\def\R{\,{\rm I\kern-.15em R}}
\def\N{\,{\rm I\kern-.15em N}}
\def\N{\,{\rm /\kern-.15em R}}

\def\mhz{\,{\rm MHz}}
\def\MHz{\,{\rm MHz}}
\def\ms{\,{\rm ms}}

\def\n{\noindent}
\def\lead{\leaders\hbox to 10pt{\hfill.\hfill}\hfill}
\def\a{\"a}
\def\o{\"o}
\def\u{\"u}
\def\infinit{\infty}
\def\upr#1{\rm#1}

\def\dd{D^{\left(2\right)}}
\def\cc{C_d ^{\left(2\right)} \left(\epsilon\right)}


\newcount\glno
\def\no{\global\advance\glno by 1 \the\glno}

\newcount\secno
\def\newsec{\global\advance\secno by 1}
\def\sec{\the\secno}

\newcount\chapno
\def\newchap{\global\advance\chapno by 1}
\def\chap{\the\chapno}



\title{On the reliability of peak-flux distributions, 
with an application to solar flares}

\author{H.\ Isliker \inst{1} \and A.O.\ Benz \inst{2}}

\offprints{H.\ Isliker}

\institute{Sect. of Astrophysics, Astronomy and Mechanics \\ 
Dept. of Physics, University of Thessaloniki \\ 
540 06 Thessaloniki, GREECE \\
isliker@helios.astro.auth.gr
\and
Institute of Astronomy, ETH Zentrum \\ 
8092 Zurich, Switzerland}

\date{Received May 11, 2001; accepted June 8, 2001}

\abstract{
Narrow-band radio spikes have been recorded during a solar flare with
unprecedented resolution. This unique example allows us to study the effect
of low resolution in previously published peak-flux distributions of radio
spikes. 
We give a general, analytical expression for how an
actual peak-flux distribution is changed in shape if the
peaks are determined with low temporal and/or frequency resolution. 
It turns out that, generally, low resolution tends
to cause an exponential behavior at large flux values if
the actual distribution is of a power-law shape. 
The distribution may be severely
altered if the burst-duration depends on the peak-flux.
The derived expression is applicable also to peak-flux distributions
derived at other wavelengths (e.g.\ soft and hard X-rays, EUV). 
We show that for the analyzed spike-event
the resolution was sufficient for a reliable peak flux distribution.
It can be fitted by generalized 
power-laws or by an exponential.
}

\maketitle

\section{Introduction}

Statistical flare models envisage the flare process as an ensemble
of sub-processes and do not focus on the single constituent processes.
The most prominent of these global
models are the so-called Cellular Automaton models 
(Lu \& Hamilton 1991; 
Lu et al.\ 1993; 
Vlahos et al.\ 1995;
Georgoulis \& Vlahos 1996; 
Galsgaard 1996;
Georgoulis \& Vlahos 1998;
MacPherson \& MacKinnon 1999;
Isliker et al.\ 2000, 2001). 
They assume flares to be fragmented and stochastic 
processes, and so the need for a comparison through
statistical quantities is present.
Furthermore, there is direct evidence that flares are 
fragmented to some unknown level 
(deJager \& deJonge 1978, Benz 1985, Aschwanden et al.\ 1990), 
and that flares really are stochastic processes 
(Isliker \& Benz 1994; Isliker 1996;
Ryabov et al.\ 1997, Veronig et al.\ 2000).

Observed peak-flux distributions of flares and flare fragments
are used to test statistical flare models.
Systems in a state of self-organized criticality (SOC) lack any 
characteristic scale and thus show power-law distributions.
Stochastic growth in uncorrelated regions of instability yield
log-normal peak-flux distributions found in interplanetary
type III bursts (Cairns \& Robinson 1997). Open driven plasmas
yield burst-like pulses distributed exponentially at large fluxes
(Robinson et al.\ 1996).
 
There exists a number of observational studies of peak-flux 
distributions of flare-related (non-thermal) emissions,
in the hard X-ray range as well as in the radio range (see
references in Aschwanden et al. 1998). In this article, we
will concentrate on narrow-band, millisecond spike events
in the radio range. 
Robinson et al.\ (1996) have determined a peak-flux
distribution from a single-frequency spike observation,
which they find to be exponential for high-flux values.
Aschwanden et al. (1998) have analyzed
some spike events with poor temporal resolution
(typically one measurement point per spike event), and found exponential 
distributions.
M\'esz\'arosov\'a et al.\ (2000) analyzed single-frequency measurements
of spikes. They found exponential and power-law distributions, the
latter being very small in extent, however (much less than one decade).
Such analyses raise a question: since every measurement
is only at discrete points in time and frequency, a detected
peak in an observation is in general not identical with the
true peak which would be seen if continuous recording
were feasible. The detected peak (termed pseudo-peak in
the following) is likely to be further away from the
true peak, the lower the time- and frequency-resolutions.
Therefore the derived peak-flux distribution must
be expected to be biased. Furthermore, peak-fluxes in the
radio-range have mostly been determined at fixed frequencies, 
neglecting completely the fact that one may be far
from the true peak in the frequency-direction. The question
therefore is what bias in a peak-flux distribution must be
expected due to finite and possibly low resolution in time 
and/or frequency.

We give an analytical expression of how a given true
peak-flux distribution is changed when determined with
finite time- and frequency-resolution (Sec.\ 2).
The expression is independent of the wave-length range under
consideration, it can also be applied e.g.\ to soft and hard 
X-rays, or to EUV.  
We then present the peak-flux distribution of narrow-band radio 
spikes in an event measured with unprecedented high time and 
frequency resolution (Sec.\ 3). We will discuss the
peak-flux distribution of the spike event with the 
introduced statistical 
theory, as well as the distributions reported in the literature, 
which are subject to poor resolution, or even derived 
without frequency information. We will address the
question of how good the time and frequency resolution must
be (in terms of duration and bandwidth of the events), in
order that the detected peak-flux distributions are near to
the true ones, and what is to be expected if the frequency
information is not available or not taken into account (Sec.\ 4).

\section{The biasing of peak-flux distributions through
finite resolution}

In this section, we will establish the analytical expression
which relates the peak-flux distribution of the true peaks
to the distribution of the observed ones with
given time and frequency 
resolutions, which we will refer to as pseudo-peaks. 
We start by making the following
definitions:

\noindent
Let $a$ be the amplitude of a pulse (e.g. spike), and assume
a pulse-shape of the form
\begin{equation}
\Phi(\nu,t,a) = a\cdot\Phi_\nu(\nu\, g(a))\cdot\Phi_t(t\, h(a))
\equiv a\cdot s
\end{equation}
with $\Phi_\nu(0)=\Phi_t(0)=1$, and $g(a)$ and $h(a)$ functions of
the amplitude $a$ (for instance $g(a) =1$, $a$, or $1/a$),
and where we have introduced the abbreviation $s=s(\nu,t;a)$. 
Such a
pulse-shape is reasonably general, but of course not completely 
so: we assume that the profiles in the frequency and in the
time direction are independent, and
particularly, it is assumed that the amplitude 
causes merely a scaling of the duration and/or bandwidth, 
and that there is no dependence on an additional,
hidden, possibly stochastic parameter.

In statistical language, we consider the ideal measurement 
of the true peak-fluxes as the outcome of a random
variable $A$, with probability density $p_A(a)$ 
($a_1 \le a \le a_2$)
(which, in other words, is the normalized peak-flux distribution). 
Analogously, the measurement of pseudo peak-fluxes 
(the ones subject to finite temporal and spatial
resolution) is considered as the outcome of a random variable 
$R$, with probability density $p_R(r)$. The question is
what the relation between $p_A(a)$ and $p_R(r)$ is, i.e. we need
to find the connection between $A$ and $R$:

Measuring the pseudo-peaks can be viewed as choosing
a random point $(N,T)$ in the $\nu$-$t$-plane within a 
certain rectangle 
around the true peak, whose side-length are $\tau_t$ and $\tau_\nu$
(the temporal and spatial resolutions, respectively), and
reading out the flux-value at this point. Hence, the uniform 
probability for choosing a random point $(N, T)$ in the
rectangle of the $\nu$-$t$-plane is transformed into a probability 
distribution on the flux axis: for a given pulse
with amplitude $A=a$, the random point $(N,T)$
transforms into a random point $a \cdot S$ on the flux axis
through the pulse-shape, 
$a \cdot S \equiv a\cdot\Phi_\nu(N)\,\Phi_t(T)$, which
will give the probability distribution $p_S(s)$ of 
the random variable $S$. 
More generally, 
we assume the pulse-shape to depend on the amplitude 
of the pulse (Eq. 1), so that S is given as
\begin{equation}
S = \Phi_\nu(N\cdot g(a))\cdot \Phi_t(T\cdot h(a)).
\end{equation}
Its probability distribution is conditionally dependent 
on the amplitude, $p_S(s \vert a)$, and its range is
$s_1(a) \le S \le s_2(a) \equiv 1$ ($p_S(s \vert a)\,ds$ 
denotes the 
{\it conditional} probability, i.e.\ the 
probability for $S$ to assume values in $[s,s+ds]$, given that $A$ is known
to assume the value $a$). The random point 
on the flux-axis
$a\cdot S$ is 
the pseudo-peak flux, so that the
relation between the true and the pseudo peak-flux is, in terms of
random variables,
\begin{equation}
R=A \cdot S        
\end{equation}
and the wanted pseudo-peak flux distribution $p_R(r)$ is 
given as
\begin{equation}
p_R(r) = p_{A \cdot S}(r) .
\end{equation}

Of course, it would be of interest to invert the problem, 
i.e.\ to derive from given $p_R(r)$ and $p_S(s)$ the true
distribution $p_A(a)$. However, this is not possible, as will
be shown in Sec.\ 2.4, since $p_R(r)$ and $p_S(s)$ do not contain
enough information 
to uncover $p_A(a)$.

\medskip\noindent
Three tasks are to be worked out now:

\begin{itemize}
\item[1.] The rectangle in the $\nu$-$t$-plane, out of which a random
  point is chosen, has to be determined (to find the probability 
  density of the point $(N,T)$, which is needed to
  derive $p_S$ through Eq.\ 2).

\item[2.] The probability $p_S(s\vert a)$ has to be derived 
 (through Eq.\ 2), since it is an input to Eq.\ (4).

\item[3.] The probability distribution of the product of the random 
  variables $A$ and $S$ has to be found to evaluate Eq.\ (4), 
  which yields the wanted $p_R$.
\end{itemize}

\subsection{The rectangle around the true peak in which the
pseudo-peak is located}

The probability density $p_{N,T}(\nu,t)$ for the random point
$(N,T)$ in the $\nu$-$t$-plane, at which the pseudo-peak flux is
measured, is uniform in some region around the true peak
(this follows from the complete absence of correlations between 
the measurement and the measured). In order to
specify $p_{N,T}(\nu,t)$, the shape and size of this region have
to be determined.      
                                                             
Without loss of generality, we may assume a true peak to
occur at $t_0 = 0$, $\nu_0 = 0$. Since we assume the burst-profile
to factorize (Eq.\ 1), we can expect the measurement point
$(N,T)$ to lie in a rectangle,                    
\begin{equation}
(N,T) \in [L_{\nu-},L_{\nu+}] \times [L_{t-}, L_{t+}]
\end{equation}
around the true peak at $(t_0,\nu_0) = (0,0)$, and we may derive
the time- and the frequency-interval independently.
Note that we explicitly treat the case where the pulse-shape 
is asymmetric. If the pulse shape is symmetric, then 
the rectangle is simply given as $[-\tau_{\nu}/2,\tau_{\nu}/2] \times
[-\tau_{t}/2, \tau_{t}/2]$, 
as follows straightforwardly from symmetry considerations.

We start with treating the time-direction. Assume 
a true peak to be located at
$t_0 = 0$. In the measurement procedure, a grid of 
time-points $t_i$ is put onto the $t$-axis (with $t_i - t_{i-1} = \tau_t$, 
$\forall i$,
with $\tau_t$ the time resolution), which is randomly positioned
relative to $t_0 = 0$. At one of these points, say at $t_i$, the
measured flux will be highest, and a pseudo-peak is detected. This means 
that (i) $a\Phi_t(t_i) \ge  a\Phi_t(t_i +\tau_t)$, and (ii)
$a\Phi_t(t_i) \ge a\Phi_t(t_i - \tau_t)$ (for now we assume the burst 
profile to be independent of the amplitude $a$). Obviously, the
pseudo-peak occurance-time $t_i$ lies in the interval
\begin{eqnarray}
t_i \in  \Big\{ t^\prime \Big\vert 
                 \Phi_t(t^\prime) \ge \Phi(t^\prime + \tau_t) , 
\  &{\rm and}&\   \Phi_t(t^\prime) \ge \Phi_t(t^\prime - \tau_t) ,
\nonumber \\
\  &{\rm and}&  \vert t^\prime \vert \le  \tau_t \Big\}            
\end{eqnarray}
The left boundary $L_{t-}$
is given by the equation $\Phi_t(L_{t-}) = \Phi_t(L_{t-} + \tau_t)$, 
with allowed values of $L_{t-}$ in $[-\tau_t, 0]$, and the right boundary
$L_{t+}$
is analogously given by the equation $\Phi_t(L_{t+}) = \Phi_t(L_{t+} -
\tau_t)$
in the range $L_{t+} \in [0, \tau_t]$. In order that these solutions are
unique, we need to demand that the pulse-shape is convex 
($\Phi_t^{\prime\prime}  < 0$), i.e. strictly increasing until the peak,
and then strictly decreasing with time, which is reasonable for
a pulse-shape. If $L_{t-} + \tau_t$ is inserted into the equation for
$L_{t+}$, then it is seen that it solves this equation, whence  
\begin{equation} 
L_{t+} = L_{t-} + \tau_t
\end{equation}
--- the size of the interval out of which a random point   
is chosen in a measurement is $\tau_t$, the time-resolution. Of
practical interest in the following will be that
\begin{equation}
\Phi_t(L_{t-}) = \Phi_t(L_{t+}) .                                        
\end{equation}

So far, we have omitted the scaling factors $h(a)$ and
$g(a)$. $L_{t-}$, for instance, would actually have to be determined 
by the equation 
$\Phi_t(L_{t-} h(a)) = \Phi_t ((L_{t-} - \tau_t) h(a) )$ 
instead of the one above, and in general one expects the solution 
$L_{t-}$ to depend on $a$. Under quite general assumptions,
however, $L_{t-}$ is independent of $a$, e.g. for Gaussian, exponential, 
or power-law pulse-shapes. All these examples
are pulse-shapes with the general form 
$a\Phi_{t-} = a \phi_0(c + \phi_-(t\,h(a)))$ for the left branch, and 
$a\Phi_{t+} = a \phi_0(c + \phi_+(t\,h(a)))$
for the right branch, where $\phi_0$ is any invertible function
(e.g.\ an exponential), $c$ a constant, and $\phi_-$ and $\phi_+$ may  
be different, but homogeneous of the same degree (i.e.\ 
$\phi_i(bt) = b^\kappa \phi_i(t)$, for $i = -,+$, and $\kappa$ a constant),
e.g.\ a
power-law. 

Completely analogously, $L_{\nu+}$ and $L_{\nu-}$ are determined,  
and again $L_{\nu+} = L_{\nu-} + \tau_\nu$ ($\tau_\nu$ denotes the
frequency 
resolution) and $\Phi_\nu(L_{\nu-}) = \Phi_\nu(L_{\nu+})$ hold. 
With the determination of the rectangle in which the random point $(N,T)$ 
lies, the probability density $p_{N,T}$ follows immediately as
\begin{equation}
 p_{N,T}(\nu,t) = p_N (\nu) p_T (t) = {1 \over  \tau_t \tau_\nu},
\end{equation}
since it is uniform, as explained at the beginning of this subsection.

\subsection{The determination of $P_S(s\vert a)$}

Instead of the probability density $p_S(s\vert a)$, we will 
derive the cumulative probability distribution 
$P_S(s\vert a) := \int_{s_1}^s p_S(s^\prime \vert a)\,ds^\prime$, since in
general 
the density has a singularity
at the peak ($s=1$) of the pulse. $P_S(s\vert a)$ is given implicitly
by Eq.\ (2) as a function of the pulse-shape.
Starting from the definition of $P_S(s\vert a)$ and inserting
Eq.\ (2), we have
\begin{eqnarray} 
P_S(s\vert a) 
      &\equiv& {\rm prob}\Big[S \le s \,\Big\vert\, 
                            {\rm given\ that\ } A=a \Big]   \nonumber \\
      &=& {\rm prob}\Big[\Phi_\nu(N \cdot g(a)) \cdot \Phi_t(T \cdot h(a))
\le s
                                                      \,\Big\vert\, a \Big]  
                                                                 \nonumber \\
      &=& {\rm prob}\Big[(T,N) \in  \nonumber \\
       & &  \big\{(t,\nu) \,\big\vert\, \Phi_\nu(\nu\, g(a))\cdot 
                                              \Phi_t(t\, h(a)) \le s,\ {\rm and}
                                                                 \nonumber \\
      & & \ L_{\nu-} \le \nu \le L_{\nu+},\ 
                             {\rm and}\  L_{t-} \le t \le L_{t+} \big\} 
                                                          \,\Big\vert\, a \Big]
                                                                 \nonumber \\
    &=& \int\!\!\!\!\!\!\!\!\!\!\!\!\!\!\!\!\!\!\!\!\!\!\!\!\!\!\!\!\!\!\!\!
  \int\limits_{\begin{array}[t]{c}\Phi_\nu (\nu \cdot g(a))  \, 
                                                           \Phi_t(t \cdot h(a)) \le s \\
                     L_{\nu-} \le \nu \le L_{\nu+} \\
                      L_{t-} \le t \le L_{t+} \end{array}}
            \!\!\!\!\!\!\!\!\!\!\!\!\!\!\!\! \!\!\!\!\!\!\!\!\!\!\!\!\!\!\!\!
              p_T(t) \, p_N(\nu) \,dt \,d\nu 
\end{eqnarray}
where $p_T(t)p_N(v)$ is given by Eq.\ (9).                      

Because of the possible asymmetry of the pulse-shape,
it is necessary to treat the four quadrants separately. In        
each quadrant then, the integration limits in Eq. (10) imply 
four different cases, depending on the value of $s$. If in
the four quadrants we denote the respective inverses of the
pulse-shape by $\Phi_{t+}^{-1}$, $\Phi_{t-}^{-1}$, and $\Phi_{\nu+}^{-1}$, 
$\Phi_{\nu-}^{-1}$, and if we write
$L_t$ where one can insert either $L_{t+}$ or $L_{t-}$ without 
changing the numerical values of the respective expressions (due
to Eq.\ 8), and analogously $L_\nu$, then, in the first quadrant,
we have (a substitution $\bar t := \Phi_\nu(\nu)\Phi_t(t)$ allows to 
calculate
the $t$-integral in Eq.\ 10):

\begin{itemize}
\item[0.] for $s \leq \Phi_\nu(L_{\nu}\, g(a))\,\Phi_t(L_{t}\, h(a))$
\begin{equation}
P_S(s\,\vert\, a) = 0
\end{equation}


\item[I.]\ for
$\,\Phi_\nu(L_{\nu}\, g(a)) \leq s$, and 
$\,\Phi_t(L_{t}\, h(a)) \leq s$, and $s\le 1$
\begin{eqnarray}
P_S(s\,\vert\, a) 
              &  = &  \ {\vert L_{\nu+}L_{t+} \vert \over \tau_\nu \tau_t}
                                          \nonumber \\
                 - &{1\over \tau_\nu \tau_t}&\,{1\over g(a)}{1\over h(a)}
                  \left\vert  \int\limits_{0}
                               ^{\Phi_{\nu+}^{-1}(s)}
\!\!\!              d\nu\, \Phi_{t+}^{-1}\left({s\over \Phi_\nu
(\nu)}\right)\right\vert
\end{eqnarray}

\item[II.]\ for
$\Phi_\nu(L_{\nu}\, g(a)) \leq s \leq \Phi_t(L_{t}\, h(a)) $
\begin{eqnarray}
P_S(s\,\vert\, a) 
        &    = &    {\vert L_{\nu+}L_{t+}\vert \over \tau_\nu \tau_t}    
\nonumber \\
               -& {1\over \tau_\nu \tau_t} & \,{1\over g(a)}{1\over h(a)}
              \left\vert     \int\limits_{\Phi_{\nu+}^{-1}\left( {s\over 
                                           \Phi_t(L_{t}\, h(a))}\right)}
                              ^{\Phi_{\nu+}^{-1}(s)}
\!\!\!\!\!\!\!\!\!\!\!\!\!\!\!
              d\nu\, \Phi_{t+}^{-1}\left({s\over \Phi_\nu (\nu)}\right)
\right\vert
                                                  \nonumber \\
              & -& \left\vert {L_{t+} \over \tau_\nu \tau_t} \, 
                 \Phi_{\nu+}^{-1}\left( {s\over \Phi_t(L_{t}\, h(a))}\right)
                  \, {1\over g(a)}  \right\vert                
\end{eqnarray}

\item[III.]\ for
$\Phi_t(L_{t}\, h(a)) \leq s \leq \Phi_\nu(L_{\nu}\, g(a))$ 
\begin{eqnarray}
P_S(s\,\vert\, a) 
               & = &  {\vert L_{\nu+}L_{t+}\vert \over \tau_\nu \tau_t}   
\nonumber \\
                  -& {1\over \tau_\nu \tau_t}&\,{1\over g(a)}{1\over h(a)}
             \left\vert       \int\limits_{0}
                               ^{L_{\nu+}g(a)}
\!\!\!             d\nu\, \Phi_{t+}^{-1}\left({s\over \Phi_\nu (\nu)}\right)
\right\vert
\end{eqnarray}

\item[IV.] for
$\Phi_\nu(L_{\nu}\, g(a)) \Phi_t(L_{t}\, h(a)) \le s$, and
$s \leq \Phi_\nu(L_{\nu}\, g(a))$, and 
$s \leq \Phi_t(L_{t}\, h(a))$
\begin{eqnarray}
P_S(s\,\vert\, a) 
            & = &  {\vert L_{\nu+}L_{t+}\vert \over \tau_\nu \tau_t}   
\nonumber \\
               &- {1\over \tau_\nu \tau_t}&\,{1\over g(a)}{1\over h(a)}
            \left\vert   \int\limits_{\Phi_{\nu+}^{-1}\left( {s\over 
                                                \Phi_t(L_{t}\, h(a))}\right)}
                              ^{L_{\nu+} g(a)}
\!\!\!\!\!      \!\!\!\!\!      \!\!\!\!\!      
            d\nu\, \Phi_{t+}^{-1}\left({s\over \Phi_\nu (\nu)}\right)
\right\vert
                                           \nonumber \\
               &-& \left\vert {L_{t+} \over \tau_\nu \tau_t} \, 
                 \Phi_{\nu+}^{-1}\left( {s\over \Phi_t(L_{t}\, h(a))}\right)
                  \, {1\over g(a)}  \right\vert               
\end{eqnarray}

\item[V.]\ for $1 \leq s$
\begin{equation}
P_S(s\,\vert\, a) = {\vert L_{\nu+}L_{t+}\vert \over \tau_\nu \tau_t}
\end{equation}

\end{itemize}

\noindent
The non-trivial cases are for intermediate ranges of $s$, i.e.\ 
for $\Phi_\nu(L_{\nu}g(a)) \Phi_t(L_{t} h(a)) \le s \le 1$. The formulae for
the other three quadrants are gained completely analogously to the given 
ones, by just correspondingly interchanging the 
indices $t+$, $t-$, $\nu+$, $\nu-$ (it is for this purpose
that we had to write absolute values for all the appearing
terms). If the pulse-shape is symmetric in $\nu$ and $t$, then
the contributions of the four quadrants are equal.

\subsection{The probability distribution of the product $A\cdot S$}

In the last step, we determine the probability distribution
$p_R$ of the pseudo-peaks, i.e. the probability distribution
 of the product $A \cdot S$ (Eqs.\ 3 and 4). The case of multiplying two
random variables which are independent and have infinite
range is found in standard textbooks. However, since both,
$A$ and $S$, have finite range, and since $S$ is conditionally
dependent on $A$, it is worthwhile to give the respective
formulae. Essentially, the probability distribution of the
product $A \cdot S$ equals $p_{A,S}(a, s)$, integrated over the 
subregion of the region $[a_1, a_2] \times [s_1(a), s_2(a)]$ 
where $as \le r$ is
fulfilled ($p_{A,S}(a, s)$ is the joint probability distribution of
$A$ and $S$). Using that $p_{A,S}(a,s) = p_A(a)p_S(s\vert a)$, we 
explicitly have
\begin{eqnarray}
P_R(r) &\equiv& {\rm prob}[R \le r] = {\rm prob}[A\cdot S \le r]  \nonumber
\\
        &=& \int\!\!\!\!\!\! \!\!\!\!\!\! \!\!\!\!\!\! \!\!\!\!\!\!
            \int\limits_{ \begin{array}[t]{c}  as \le r \\
                                      a_1 \le a \le a_2   \\
                                  s_1(a) \le s \le s_2(a) \end{array} }     
           \!\!\!\!\!\!\!\!\!\!   p_{A,S}(a,s)\, da\, ds                   \nonumber \\
        &=& \int\limits_{\begin{array}[t]{c}  a_1 \le a \le a_2       \\
                                       a \le r/s_1(a) \end{array} }          
            \!\!\!\!\!   \!\!\!\!\!     p_A(a) \, da   
          \int\limits_{\begin{array}[t]{c}  s_1(a) \le s \le s_2(a) \\
                                       s \le r/a \end{array}}
             \!\!\!\!\!\!\!\!\!\!  \!\!\!\!\!           p_S(s\,\vert\, a)\, ds    
\end{eqnarray}
where $s_1a_1 \le r \le s_2a_2$. (As mentioned in Sec.\ 2.2, 
$p_S(s\vert a)$
might have a singularity at $s =1$, and we must therefore
use the cumulative probability distribution $P_S (s\vert a)$, which 
is achieved if also for $R$ we use the cumulative probability 
distribution $P_R(r)$, as done in Eq.\ (17), and identify 
$\int_\alpha^\beta p_S(s^\prime\vert a) \, ds' 
  = P_S(\beta\vert a) - P_S(\alpha\vert a)$.)    

In the following, we write again $L_t$ if inserting $L_{t+}$  
or $L_{t-}$ does not change the respective numerical values,
and correspondingly $L_\nu$ is used instead of 
$L_{\nu+}$ or $L_{\nu-}$. 
For intermediate values of $r$, i.e.\ for 
$a_1 \cdot \Phi_\nu(L_\nu g(a_1)) \Phi_t(L_t h(a_1)) \le r \le a_2$, 
four non-trivial 
cases of integration limits turn out to exist, depending on the 
values of r. We have:

\begin{itemize}

\item[0.]\ for
$r\leq a_1\cdot\Phi_\nu(L_{\nu}\, g(a_1))\,\Phi_t(L_{t}\, h(a_1))$
\begin{equation}
P_R(r) = 0
\end{equation}

\item[I.]\ for 
$r\leq a_2\cdot\Phi_\nu(L_{\nu}\, g(a_2))\,\Phi_t(L_{t}\, h(a_2))$, and
$r\leq a_1$
\begin{equation}
P_R(r) =  \!\!\!\!\! \!\!\!\!\! \int\limits_{a_1}
                   ^{a\leq r/\Phi_\nu(L_{\nu}\, g(a))\,\Phi_t(L_{t}\, h(a))} 
\!\!\!\!\! \!\!\!\!\! \!\!\!\!\!          da\,p_A(a)\,P_S(r/a\,\vert\, a)
\end{equation}

\item[II.]\ for 
$a_1\leq r \leq a_2\cdot\Phi_\nu(L_{\nu}\, g(a_2))\,\Phi_t(L_{t}\,
h(a_2))$
\begin{equation}
P_R(r) = P_A(r) \, + \!\!\!\!\! \!\!\!\!\! \!\!\!\!\!
         \int\limits_{r}
                   ^{a\leq r/\Phi_\nu(L_{\nu}\, g(a))\,\Phi_t(L_{t}\, h(a))} 
\!\!\!\!\!  \!\!\!\!\! \!\!\!\!\! \!\!\!\!\!         da\,p_A(a)\,P_S(r/a\,\vert\, a)
\end{equation}

\item[III.]\ for
$a_2\cdot\Phi_\nu(L_{\nu}\, g(a_2))\,\Phi_t(L_{t}\, h(a_2)) \leq r\leq
a_1$
\begin{equation}
P_R(r) = \int\limits_{a_1}
                    ^{a_2} 
          da\,p_A(a)\,P_S(r/a\,\vert\, a)
\end{equation}

\item[IV.]\ for
$a_2\cdot\Phi_\nu(L_{\nu}\, g(a_2))\,\Phi_t(L_{t}\, h(a_2)) \leq r$, and
$a_1\leq r$, and $r\le a_2$
\begin{equation}
P_R(r) = P_A(r) + 
         \int\limits_{r}
                    ^{a_2} 
          da\,p_A(a)\,P_S(r/a\,\vert\, a)
\end{equation}

\item[V.]\ for $a_2 \leq r$
\begin{equation}
P_R(r) = 1
\end{equation}

\end{itemize}
Inserting $P_S(s\vert a)$ (Eqs.\ (11) to (16)) into the formulae 
for $P_R(r)$ yields the desired pseudo peak-flux distribution.

\subsection{Inversion is not possible}

To derive the true peak flux distribution from a given
pseudo-peak flux distribution (the inverse problem), we
have to proceed as follows: from Eq.\ (3) we find
\begin{equation}
A = R/S,
\end{equation}
and we can uncover the true distribution from the pseudo
one analogously as we had proceeded in Eq. (17):
\begin{eqnarray}
P_A(a) &\equiv& {\rm prob}[A \le a] = {\rm prob}[R/S \le a]  \nonumber \\
      &=& \int\!\!\!\!\!\!\int\limits_{r/s \le a}  p_{R,S}(r,s) \,dr\,ds
                                       \nonumber \\
      &=&   \int\!\!\!\!\!\!\int\limits_{r/s \le a}  p_R(r\vert s)p_S(s)\,dr\,ds       
\end{eqnarray}
so that we need to know the conditional probability $p_R(r\vert s)$
of $R$ given that $S$ is known. That $R$ is not independent of
$S$ is evident from the fact that $R$ is equivalent to $A\cdot S$,
and one explicitly finds, if $p_A(a)$ is assumed to be known,
that
\begin{equation}
P_R(r\vert s) = P_A(r/s)                            
\end{equation}

What is measured, however, is
\begin{equation}
p_R(r) =   \int\limits_{s_1\leq s \le s_2}   p_R(r\vert s)p_S(s) \,ds , 
\end{equation}
the distribution of $R$ irrespective of the value of $S$, so that
all the conditional information is lost. In other words, to
uncover $P_A(a)$ one has to know $P_R(r\vert s)$, which is 
essentially equivalent to knowing the true peak-flux distribution
$P_A(a)$ (see Eq.\ 26), and which would be feasible only in
continuous observations. Whence it follows that uncovering 
the true distribution from the measured one is not
possible.

\section{The peak-flux distribution of solar
narrow-band spikes}

\begin{table}

\caption[]{Fits of different functional forms to the peak-flux
distributions of the event 1982/06/04 (see Sec.\ 3), where the
four data-sets are distributions derived from a) peaks in 2D
plane,  $\tau_t = 2\ms$, $\tau_\nu = 1\MHz$ (data-set a); b) peaks at 
$\nu = 362\MHz$, $\tau_t = 2\ms$ (data-set b); c) peaks in 2D plane,
$\tau_t= 100\ms$, $\tau_\nu = 1\MHz$ (data-set c); d) peaks at $\nu =
362\MHz$,
$\tau_t = 100\ms$ (data-set d). If the $\chi^2$ approved a fit, then the
power-law index or an 'o.k.' is stated, else a '---' is noted.}

\begin{tabular}{|c|c|r|r|r|c|}   \hline
 data &  nr.\ of &  \multicolumn{1}{c|}{$ax^c$} & \multicolumn{1}{c|}{$a(x -
b)^c$} &  
                              \multicolumn{1}{c|}{$a(x - b)^c + d$} &  $ae^{cx}$   \\
 set  &  peaks  &  \multicolumn{1}{c|}{$c =$}   &    
                   \multicolumn{1}{c|}{$c =$}  &  \multicolumn{1}{c|}{$c =$}  &  
\\
                         \hline\hline
  a   &   59    &   ---     &  -15.6$\pm$6.8  &    -2.7$\pm$7.5  &   o.k.  \\
                           \hline
  b  &    144  &  ---    &     ---        &   -14.9$\pm$7.3   &   o.k.    \\
                            \hline
  c  &   38    &  -1.2$\pm$0.2 &  -3.5$\pm$6.9  &  -0.8$\pm$7.9 & o.k.  \\
                             \hline
  d   &   76    &    ---  &     -16.3$\pm$4.8  &  -15.8$\pm$6.8  & o.k.  \\
                                      \hline
\end{tabular}

\end{table}

The solar narrow-band millisecond spike event we analyzed 
was observed by the ETH Zurich radio-spectrometer
on 1982/06/04, 13:38:41 UT (the event is published and
described in G\"udel \& Benz, 1990;  
Csillaghy \& Benz, 1993). The resolution is 2
ms in time and 1 MHz in frequency (from 361 to 364
MHz), whereas the spikes have a typical duration of 73 ms
(FWHM; G\"udel \& Benz, 1990), and a typical bandwidth
of 7 MHz (FWHM; Csillaghy \& Benz, 1993). 
Hence the spikes are well resolved in time. Also in frequency, the spikes
are resolved although the observation range (4 MHz) is smaller than the
typical bandwidth, as only spikes with peaks unambiguously in the range
are taken into account.
The peaks
were determined by looking for strong enough local maxima above the
noise-level, and a constant background was
subtracted (representing the quiet Sun). The normalized
distribution of the respective peak-fluxes is shown in 
Fig.\ 1 (solid line, with error-bars). 

Different curves were fitted to this distribution, 
and a $\chi^2$-test was performed to check
whether the fits are compatible with the data or not. The
fitted curves are: a simple power-law ($ax^c$), two generalized 
forms of power-laws ($a(x-b)^c$, and $a(x-b)^c+d$), and
an exponential ($ae^x$). The result is presented in Fig. 2 and
summarized in Table 1 (data-set a): The peak-flux distribution 
of the event can be fitted by the generalized power-laws 
as well as by the exponential, but not by the simple
power-law. The indices of the generalized power-laws 
are subject to large
errors (determined by the bootstrap resampling method)
due to the large error-bars 
(the number of peaks is relatively small) and due to the relative 
flexibility of the generalized power-laws (three resp.\ four free 
parameters and 10 data-points).


\begin{figure}
\resizebox{\hsize}{!}{\includegraphics{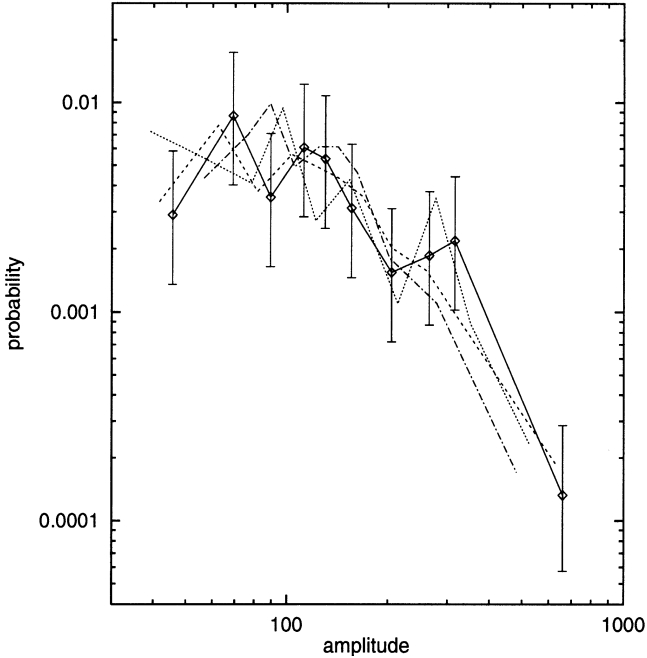}}
\caption{Peak-flux distributions (probability densities) of the
narrow-band spike event 1982/06/04 (see Sec.\ 3): a) peaks in
2D plane, $\tau_t = 2\ms$, $\tau_\nu = 1\MHz$ (solid line, with error-
bars);
b) peaks at $\nu = 362\MHz$, $\tau_t = 2\ms$ (dashed line); c) peaks in
2D plane, $\tau_t = 100\ms$, $\tau_\nu = 1\MHz$ (dotted line); d) peaks at
$\nu = 362\MHz$, $\tau_t = 100\ms$ (dash-dotted line). (The histograms
are drawn by connecting the midpoints of the bins, and the
bin-widths are such that each contains the same number
of data-points. The amplitude is in SFU.)}
\label{}
\end{figure}

\begin{figure}
\resizebox{\hsize}{!}{\includegraphics{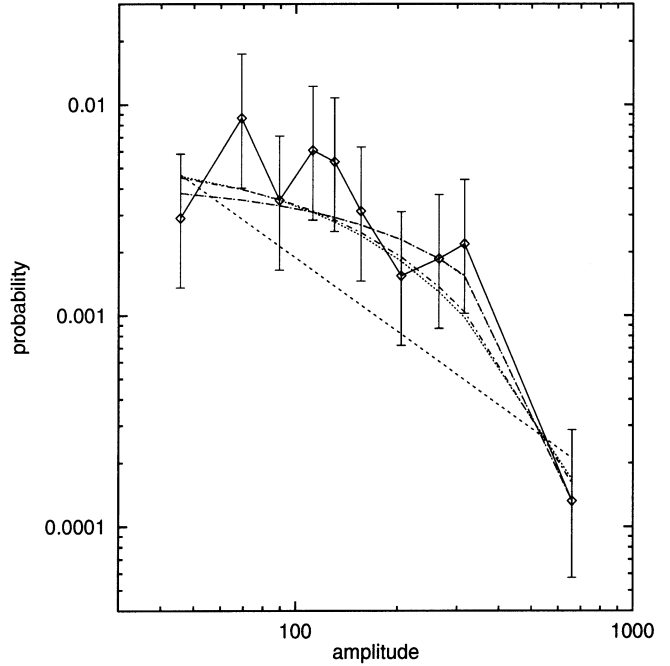}}
\caption{Fits to the peak-flux distribution (probability density)
of the narrow-band spike event 1982/06/04 (see Sec.\ 3, and also
Table 1). The peaks are determined in the 2D plane, with $\tau_t =  
2\ms$ and $\tau_\nu = 1\MHz$. Plotted are the original data (solid line),
and the fits $ax^c$ (dashed), $a(x-b)^c$ (dotted), $a(x-b)^c+d$ 
(dash-dotted), $ae^x$ (wide dash-dotted). (The histogram is drawn and
generated as described in Fig.\ 1, and the amplitude units are again SFU.)}
\label{}
\end{figure}

\section{Discussion of the empirical peak-flux
distributions}

\begin{figure}
\resizebox{\hsize}{!}{\includegraphics{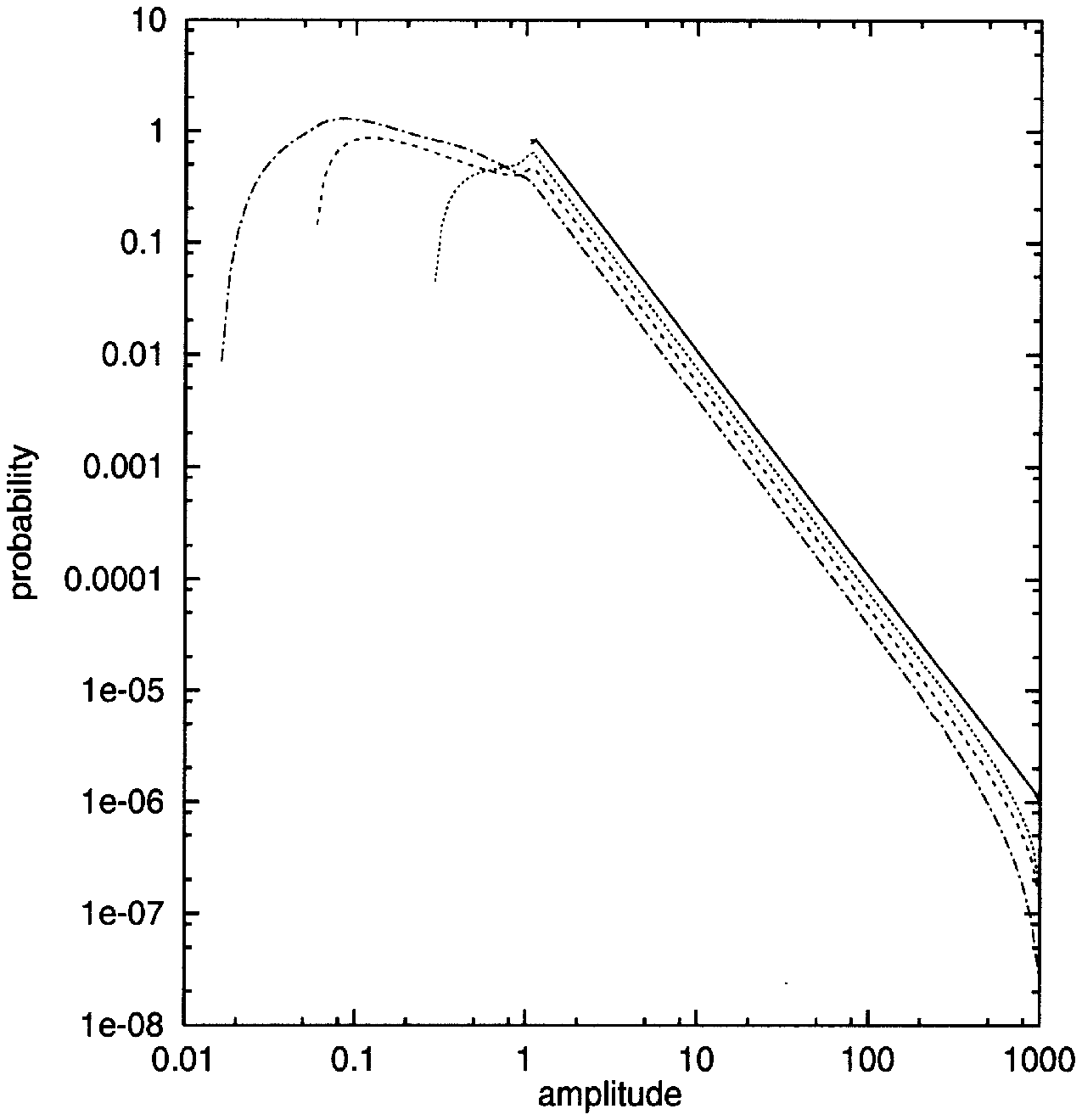}}
\caption{{\it Amplitude-independent pulse-shape:}
(0) True peak-flux distribution ($\alpha = 2$, $a_1 = 1$, $g(a) =
 h(a) \equiv 1$)  (solid),  together  with  the  pseudo peak-flux
 distributions for 
(i) $\tau_t = 2\ms$, $\tau_\nu = 1\MHz$ (dashes); 
(ii) $\tau_t= 2\ms$, $\tau_\nu = 15\MHz$ (small dashes); 
(iii) $\tau_t = 100\ms$, $\tau_\nu =1\MHz$ (dots); 
(iv) $\tau_t = 100\ms$, $\tau_\nu = 15\MHz$ (dot-dashed).
The cases (0) and (i) practically coincide.}
\label{}
\end{figure}

\begin{figure}                             
\resizebox{\hsize}{!}{\includegraphics{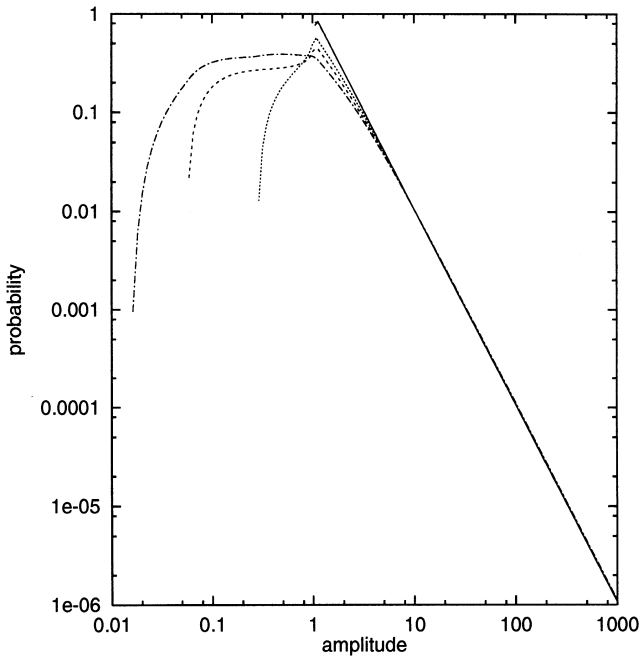}}
\caption{{\it Amplitude-dependent pulse-shape I:}
True  peak-flux  distribution  ($\alpha = 2$,  $a_1=1$,
$g(a) = h(a) = a^{-1}$) (solid), together with pseudo peak-flux
distributions for $\tau_t$ and $\tau_\nu$ as described in the caption to 
Fig.\ 3.
The cases (0) and (i) practically coincide.}
\label{}
\end{figure}

\begin{figure}
\resizebox{\hsize}{!}{\includegraphics{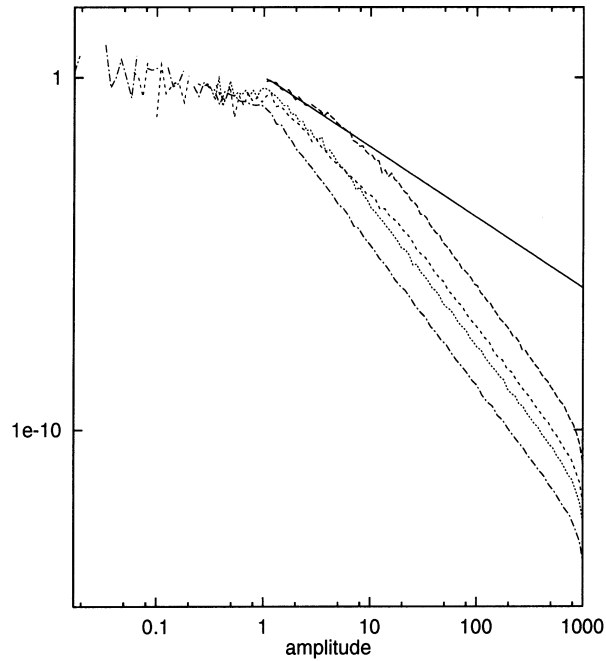}}
\caption{{\it Amplitude-dependent pulse-shape II:}
True   peak-flux  distribution  ($\alpha = 2$,  $a_1  = 1$,
$g(a) = h(a) = a^1$) (solid), together with pseudo peak-flux 
distributions for $\tau_t$ and $\tau_\nu$ as described in the caption 
to Fig.\ 3.}
\label{}
\end{figure}

\begin{figure}
\resizebox{\hsize}{!}{\includegraphics{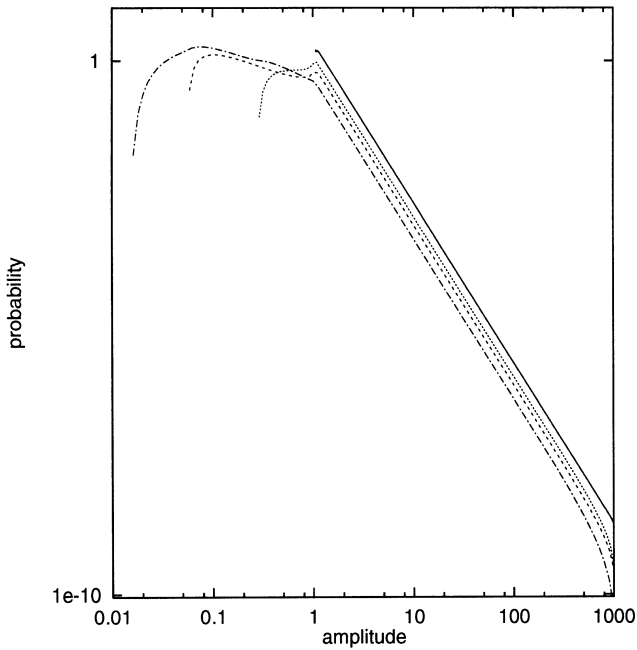}}
\caption{{\it Steeper true distribution:}
True peak-flux distribution ($\alpha = 3$, $a_1  = 1$, $g(a) =
h(a) \equiv 1$) (solid), together with pseudo peak-flux distributions
for $\tau_t$ and $\tau_\nu$ as described in the caption to Fig.\ 3.
The cases (0) and (i) practically coincide.}
\label{}
\end{figure}
 
\begin{figure}
\resizebox{\hsize}{!}{\includegraphics{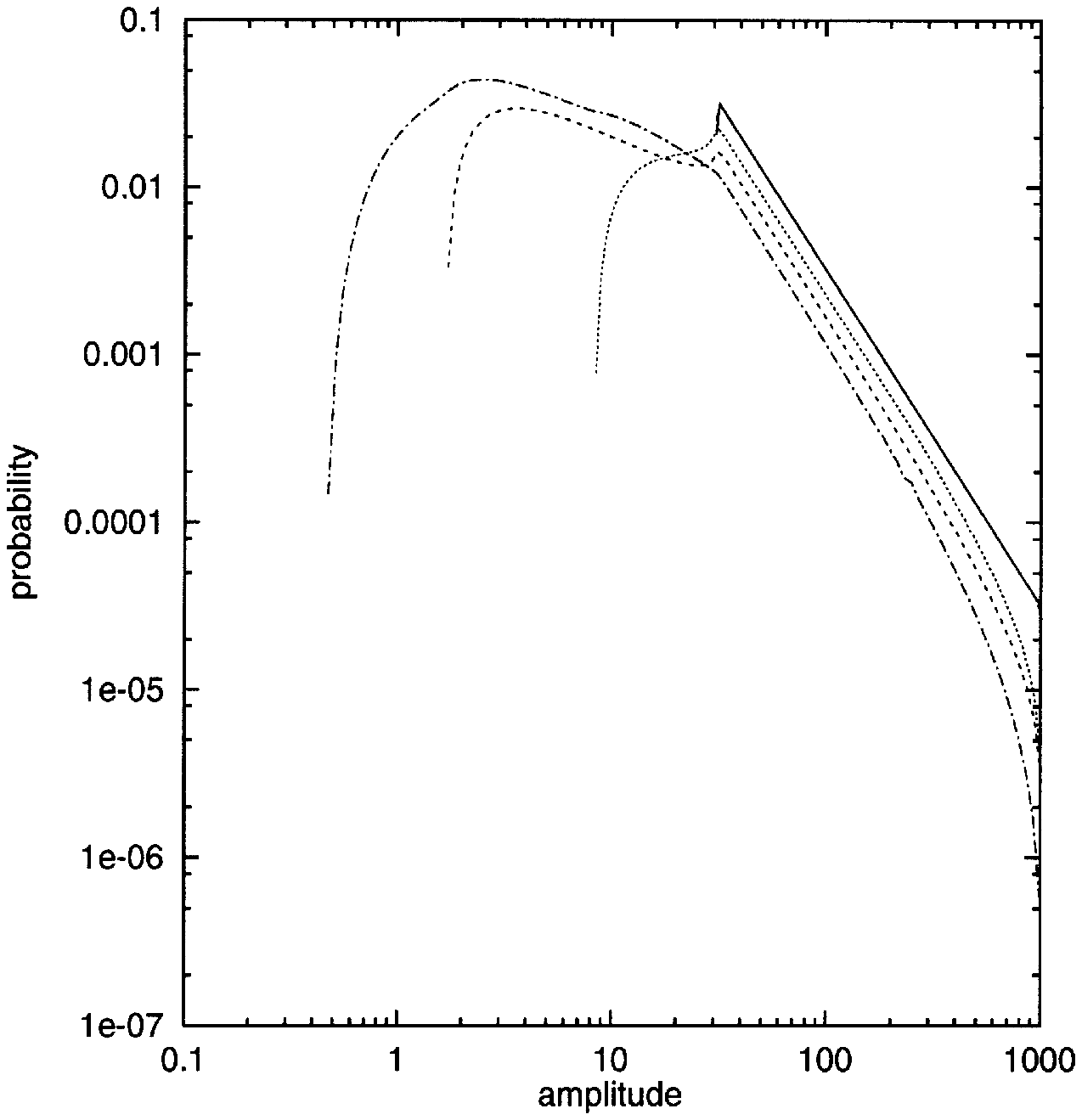}}
\caption{{\it Smaller range of amplitudes:}
True peak-flux distribution ($\alpha = 2$, $a_1 = 30$, 
$g(a) = h(a) \equiv 1$) (solid), together with pseudo peak-flux
 distributions 
for $\tau_t$ and $\tau_\nu$ as described in the caption to Fig.\ 3.
The cases (0) and (i) practically coincide.}
\label{}
\end{figure}

To apply the statistical theory introduced in 
Sec.\ 2 to the narrow-band spike event analyzed in Sec.\ 3,
we have to make an assumption about the pulse-shape of the
individual spikes. According to G\"udel \& Benz (1990),
and Csillaghy \& Benz (1993), it is reasonable to assume
a Gaussian pulse-shape, i.e.
\begin{equation}
\Phi_\nu(\nu) = e^{-{1\over 2}\left({\nu\over b_\nu}\right)^2}
\end{equation}                                           
and
\begin{equation}
\Phi_t(t) = e^{-{1\over 2}\left({t\over b_t}\right)^2}
\end{equation}                                           
with $b_\nu = 3.1\MHz$, and $b_t = 31\ms$ (implying a FWHM of
$7.3\MHz$, and $73\ms$, respectively, as reported by G\"udel
and Benz (1990), and Csillaghy and Benz (1993) for the
given event). Furthermore, we have to assume a distribution 
of the true peak-fluxes $A$: we let $p_A(a)$ be a straight
power-law
\begin{equation}
p_A(a) = C a^{-\alpha},\ \ \      a_1 \le a \le a_2
\end{equation}
with $a_2 =1000$ (from Fig.\ 1). Other true distributions can
be expected to produce qualitatively the same effects as
reported below for the case of this power-law.

The crucial parameters are the time-resolution $\tau_t$ and
the frequency-resolution $\tau_\nu$, they determine how near the
pseudo-peaks are to the true-peaks, on average. Whence,
in the following parametric study (Figs.\ 3 to 7), we 
always show the true peak-flux distribution $p_A(a)$, together 
with the pseudo peak-flux distributions for four cases of
time- and frequency-resolution: 
(i) good resolution in time and frequency, 
(ii) good resolution in time and a bad one in frequency,
(iii) bad resolution in time and a good one in frequency, and 
(iv) bad resolution in both time and frequency 
(by 'good' we mean $\tau <<$ FWHM, and 'bad'
means $\tau = 2$ FWHM). The cases (ii) and (iv) represent also
the case of single-frequency observations.

First, we investigate the case of an amplitude-independent
pulse-shape, i.e. $g(a) = h(a) \equiv 1$ (see Eq.\ 1), and we 
set $a_1 =  1$, $\alpha = 2$. Fig.\ 3 shows that for 
good resolution both in frequency and time the true and the pseudo
peak-flux distributions practically coincide. If one or both
resolutions are low, then the pseudo-peak flux
distribution is generally near the true one, with a faster 
fall-off at high flux-values, however, 
i.e. a turning from power-law to exponential behaviour.
At pseudo peak-flux values smaller than $a_1$, a completely 
artificial, relatively flat extension of the distribution appears.
Turn-overs at small fluxes in distributions detected with
low resolution(s) might thus be just the effect of an intrinsic 
low-amplitude cut-off in the true fluxes. Only high resolution
analysis can tell whether  or not such a flattening is real or not.
The turning-over effect increases if the resolution
decreases. 

We turn now to the case of amplitude dependent pulse-shapes.
If we let $g(a) = h(a) = a^{-1}$, then we get a
deviation from the power-law behavior at small flux values, 
but nice coincidence for intermediate and high values
(Fig.\ 4). For $g(a) = h(a) = a^1$ the slope (and shape at 
high flux-values) is drastically changed to a stronger fall-off 
(Fig.\ 5), the observed distribution is different even for 
good time and frequency resolution.

The general behavior demonstrated so far is rather 
independent of the shape of
the original, true peak-flux distribution: In Fig.\ 6, we let again
$g(a) = h(a) \equiv 1$, but now the power-law index $\alpha = 3$. The
corresponding 
pseudo peak-flux distributions behave analogously to the
case $\alpha = 2$ (Fig.\ 3). 

It is worthwhile noting that the parameter 
$a_1$ may enhance the effect of turning to exponential 
behavior at high flux values: in Fig.\ 7, we let
$g(a) = h(a) \equiv 1$ and $\alpha = 2$, as in Fig. 3, but $a_1 = 30$ (cf.\ 
the lower cut-off in Fig.\ 1), and the tendency seen in 
Fig.\ 3 is enhanced, now. Thus, we conclude that the roll-over 
at large amplitude is more serious for a small range of amplitudes.

Obviously, it is crucial to know whether the durations
and bandwidths of the pulses depend on the amplitude:
Csillaghy and Benz (1993) report that the bandwidth of spikes 
sometimes 
is correlated with the amplitude, sometimes it is not,
and sometimes it is anti-correlated. There is, however, no
generally holding strong tendency, so that we may assume
that the peak-flux distribution reported in Sec.\ 3 is near
the true one. 

This is also confirmed by the following: we   
artificially worsened the time and the frequency resolution 
in the data of Sec.\ 3 and compared the respective     
histograms: The dashed line in Fig.\ 1 is the histogram of   
the pseudo-peak fluxes resulting if the frequency information 
is completely neglected, i.e.\ one of the frequencies is
selected, and the peaks are determined as maxima in 
time-direction only. The dotted line is the distribution for the
resampled observation, using only every 50th data-point
in time, which yields a spectrogram with $0.1 \second$ time resolution, 
and keeping full frequency resolution. Finally, the   
dashed-dotted line is the distribution for the resampled       
data, with again fixing a frequency and neglecting completely 
the corresponding information. Obviously, 
the biasing effects are smaller than the statistical errors
in the  distributions, all four
distributions coincide within the error-bars, and the results are
fairly independent of the sampling --- only the fits seem to  
show different kinds of behavior (Table 1), but, as mentioned, 
the indices of the generalized power-laws are subject to 
large errors.
The relative robustness (within the statistical errors) 
of the distribution on under-sampling is in turn 
a signature for amplitude-independent pulse-shapes.

\section{Conclusion}

The statistical theory we introduced allows us to predict the
deviation of a pseudo-peak flux distribution from the true
peak-flux distribution if the time and frequency resolutions 
are known. It turns out that in general there is a 
tendency towards exponential behavior at large flux values; 
the more expressed, the lower the resolutions are, including in 
particular the case of single-frequency observations.
Only with high resolutions in both frequency and time 
(compared e.g.\ to the respective FWHM) are
the detected distributions reliable in the whole range. 
The dependence of the pulse-shape on the amplitude (peak-flux) is 
crucial:
If the width of a pulse (duration or bandwidth) 
is proportional to the inverse of the
amplitude, then a strong deviation from the true peak flux
distribution will result, the distributions will be steepened and 
completely biased in the whole range, even for high resolution
in both frequency and time. If the width of the pulse 
is directly proportional to the amplitude, then a bias 
(flattening) appears only in the low amplitude range.
All the biasing effects get stronger with a smaller extent of a 
distribution.

A different possible cause for a strong bias at low amplitudes 
(appearance of a relatively flat part in the detected distribution) 
is a possible
intrinsic lower cut-off of the true amplitudes. Only high resolution
data analysis can make sure whether a flattening at low amplitudes
is real or an artifact.

The example of a narrow-band spike event we analyzed is a 
unique observation with respect to the high temporal and frequency 
resolutions; the spikes are completely resolved in frequency and time.
Since moreover the burst-width seems to depend at most slightly 
on the amplitude, we conclude on the basis of our analytical study
that the peak-flux distribution we get is reliable.
It  
can be fitted by generalized power-laws or by an exponential, but
definitely not by a simple power-law. 

From our analysis, it follows that the peak-flux distributions 
of spikes reported by
Robinson et al.\ (1996),
Aschwanden et al. (1998),
and M\'esz\'arosov\'a et al.\ (2000),
though observed with poor time-resolution or without any
frequency information, are near the
true distributions, except for the high-flux part, which must be expected
to be too steep. 

The exponential distribution we find supports the open
driven plasma model (Robinson et al.\ 1996). We did though not
try to fit a log-normal distribution, which in view of the relatively 
large statistical error in the empirical distribution, 
is likely to be also accepted by the $\chi^2$-test, so that    
we cannot exclude the stochastic growth theory (Cairns \& Robinson
1997).

Conclusions on whether the distribution we find is compatible
or not with SOC models are more difficult to draw. SOC models
are models for the primary energy release, so far they do not 
include a mechanism for radio- (plasma-) emission. Moreover,
the peak-fluxes of SOC models,
which have been analyzed statistically, are defined as the
peak-fluxes of entire flares, whereas here and in the cited 
articles on radio observations, the peak-fluxes of all the 
flare-fragments are analyzed. For both reasons, a direct 
comparison of SOC models to radio data 
is not possible without substantial new developments.

The statistical theory introduced here can be applied 
to the peak-flux distributions of all the different kinds of bursts,
independent of the wavelength at which they occur,
as soon as the pulse-shape and the functional dependency of 
the pulse-shape 
on the amplitude are (at least approximately) known: 
to radio-bursts (recent empirical studies include 
Aschwanden et al.\ (1998; type III, decimetric pulsations),
Mercier and Trottet (1997; type I)), 
to soft X-rays, EUV, hard X-rays, etc. 
(see e.g.\ the review of Crosby et al.\ 1993;
Krucker \& Benz 1998).

\begin{acknowledgements} 
We thank A.\ Anastasiadis and K.\ Tsiganis for many clarifying 
discussions. The work of H.\ Isliker was 
supported by a grant of the Swiss National Science Foundation
(NF grant nr.\ 8220-046504). 
The construction of the radio spectrometers at ETH is financially
supported by the Swiss National Science Foundation 
grant nr.\ 2000-061559.
\end{acknowledgements}

\end{document}